
\magnification=1200
\overfullrule=0pt
\baselineskip=20pt
\parskip=0pt
\def\dag{\dagger}
\def\del{\partial}

\def\a{\alpha}     
\def\b{\beta}      
     
\def\d{\delta}     \def\D{\Delta}
   
\def\z{\zeta}

    \def\L{\Lambda}
	   
\def\n{\nu}        
        
\def\o{\omega}          
\def\p{\pi}        \def\P{\Pi}
\def\r{\rho}

\def\w{\omega}

\def\br{\langle}
\def\ke{\rangle}
\def\ve{\vert}

\def\Winf{$W_{\infty}\  $}
\def\kvecz{\ve z_1  \cdots z_N \ke}

\def\zbar{\bar{z}}
\def\wbar{\bar{w}}

{\settabs 5 \columns
\+&&&&SU-4240-580 \cr
\+&&&&April 1994\cr}
\bigskip
\centerline{\bf \Winf ALGEBRAS IN THE QUANTUM HALL EFFECT }

\vskip 0.4in
\centerline{ Dimitra Karabali {\footnote{*} {e-mail
address: karabali@suhep.phy.syr.edu}}}
\centerline{ Physics Department, Syracuse University, Syracuse, NY 13244
{\footnote{$^{\dag}$} {Address after September 1, 1994 : The Institute for
Advanced Study, School of Natural Sciences, Princeton, NJ 08540}}}

\vskip 0.8 in
\centerline{\bf Abstract}
\bigskip

We show that a large class of incompressible quantum Hall states correspond to
different representations of the \Winf algebra by explicit construction of the
second quantized generators of the algebra in terms of fermion and vortex
operators. These are parametrized by a set of integers which
are related to the filling fraction. The class of states we consider
includes multilayer Hall states and the states proposed by Jain to explain the
hierarchical filling fractions. The corresponding second quantized order
parameters are also given.

\pageno = 0

\vfill\eject

\noindent{\bf 1. Introduction}

The quantum Hall effect $^{[1,2]}$(QHE) appears in two-dimensional
systems of electrons in the
presence of a strong perpendicular uniform magnetic field $B$.
It is characterized by the existence of a series of plateaux where the
Hall conductivity is quantized and the longitudinal conductivity
vanishes. The Hall conductivity is
proportional to the filling fraction $\nu$, the ratio between the number of
electrons and the degeneracy of the Landau levels.
This generic feature which appears in both the integer (IQHE) and the
fractional (FQHE) quantum Hall effect is attributed to the existence of a gap,
which gives rise to an
incompressible ground state. These correspond to incompressible droplet
configurations of uniform density $\r = \n
B/2 \p$.

Recently the notion of incompressibility has been related to the existence of
an infinite dimensional algebraic structure, the \Winf algebra, which emerges
quite
naturally at least in the case of the IQHE $^{[3,4]}$.
In this case ($\nu =n$) the energy gap
is the cyclotron energy separating adjacent Landau levels and the phenomenon
can be understood in terms of noninteracting fermions. The underlying \Winf
algebra emerges as the algebra of
unitary transformations which preserve
the particle number at each Landau level and it plays the role of a spectrum
generating
algebra. The corresponding ground state satisfies highest weight
conditions$^{[4]}$.

The noninteracting picture is nonapplicable in the case of the FQHE, where
the repulsive Coulomb interactions
among electrons become important in producing an energy gap. Much of our
understanding of the FQHE relies on successful trial wavefunctions. For
example, in an attempt to explain the FQHE for $\nu = 1/m$, where $m$ is an
odd integer,
Laughlin proposed$^{[5]}$ a set of wavefunctions which turn out to be quite
close to the exact numerical solutions
for a large class of repulsive potentials$^{[6]}$.
Based on Laughlin wavefunctions, a hierarchy scheme$^{[6,7]}$ has
been developed to explain the other rational values of $\nu$. A somewhat
different approach has been developed by
Jain$^{[8]}$. In this approach the FQHE wavefunction is constructed by
attaching an
even number of magnetic fluxes to electrons occupying an integer number of
Landau levels.
This way the incompressibility of the IQHE wavefunctions is carried over to the
FQHE wavefunctions.

Following the example of the IQHE, attempts have been made to extend the
connection between the incompressibility of the ground state and the \Winf
algebra in the case of the FQHE$^{[4,9,10,11]}$. In ref.[10], using the
relation
between the $\n =1$ and $\n = 1/m$ ground state, we found that there exists
a second quantized expression of a \Winf algebra which plays the role of a
spectrum generating algebra for
Laughlin wavefunctions such that the Laughlin ground state satisfies highest
weight condition.
This provides a specific one-parameter family of
\Winf representations, the parameter being related to the filling fraction.

In this paper we extend these ideas to more general filling
fractions $\n$. We shall derive explicit representations, in terms of second
quantized fermion and vortex operators, of the \Winf algebra generators for the
cases of
multilayer incompressible states and the states proposed by Jain$^{[8]}$ to
explain the hierarchical filling fractions. All these ground states satisfy
highest weight conditions. We shall also derive second quantized expressions
for the corresponding order parameters.

More generally incompressible states can be thought of as highest weight states
of the \Winf algebra$^{[11]}$. Expressing the \Winf generators in terms of Fock
operators is a useful technique for constructing representations and of course
the highest weight state. It is thus possible to go beyond the particular
examples we have considered here and obtain other candidate states for
describing quantum Hall effect.

\bigskip
\noindent{\bf 2. \Winf algebras for IQHE}

The single-body Hamiltonian for spin polarized (scalar) fermions confined in a
two-dimensional plane, in the presence of a uniform magnetic field $B$
perpendicular to the plane is (in units $\hbar=c=e=1$)
$$
H = {1\over{2M}} ( {\bf \Pi } )^2 \eqno (2.1)
$$
where
$
\Pi^i = p_i-A^i ({\bf x}),
\  i= x, y $ and $\vec{\nabla} \times
\vec {A}=-B$.

We define the following two sets of independent raising and lowering operators
$$ \eqalign {
& a={1 \over \sqrt{2B}} (\P ^{x} -i \P ^{y})~~~~~~~~ a^{\dag}= {1 \over
\sqrt{2B}} (\P ^{x} +i \P ^{y}) ~~~~~~~~~~~[a,a^{\dag}]=1 \cr
& b=\sqrt{{B \over 2}}(X+iY)~~~~~~~~~~b^{\dag}=\sqrt{{B \over 2}}(X-iY)~~~~~~~
{}~~~ [b, b^{\dag}]=1 \cr} \eqno(2.2)
$$
where $X,Y$ are the guiding center coordinates
$$X=x-{1 \over B} \P^{y}~~~~~~~~~Y=y+{1 \over B} \P^x~~~~~~~~~~[X,Y]={i \over
B} \eqno(2.3) $$
The two sets of operators $a, a^{\dag}$ and $b,b^{\dag}$ are mutually
commuting: $[a,b]=[a,b^{\dag}]=0$.

The Hamiltonian can be written as
$$ H=\o (a^{\dag} a +1/2),~~~~~~~~~\o=B/m   \eqno(2.4) $$
and the angular momentum operator is given by
$ J= b^{\dag}b- a^{\dag}a  $.

Any state can be now decomposed in terms of the
basis $|n \ke \otimes |l \ke \equiv |n,l \ke$, which is a tensor product of
Fock states
defined by
$$ \eqalign {
 & a^{\dag} a |n \ke = n|n \ke ~~~~~~~~~~~~~  \br n'|n \ke = \d _{n',n} \cr
 & b^{\dag} b |l \ke  = l|l \ke ~~~~~~~~~~~~~  \br l'|l \ke = \d _{l',l} \cr}
\eqno(2.5) $$
Let us also introduce the coherent basis
representation for the $b$ oscillators
$$
b|\z \ke = \z |\z \ke  ~~~~~~~~ \br \z|\z ' \ke =e^{\bar{\z} \z '} ~~~~~~~~
\br l|\z \ke ={\z
^l \over \sqrt{l!}} \eqno(2.6) $$
where $ |\z \ke = e^{b^{\dag} \z} |0 \ke $ and $\int d^2 \z e^{-|\z |^2} |\z
\ke \br \z|
=1,~ d^2 \z = { {d Re \z d Im \z } \over \pi}$.

The fermionic operator can be similarly decomposed as
$$ \Psi (\vec {x} ,t) = \sum _{n,l=0} ^{\infty} C_{l}^{n} (t) \Psi _{n}
^{l-n}(\vec{x}) \eqno (2.7) $$
where $ \br l,n| \equiv <0| C_{l}^{n}$ and $\Psi _{n}^{l-n}(\vec{x}) \equiv
\br \vec{x}|n,l \ke $ is the one-body wavefunction of energy $\o (n+1/2)$ and
angular
momentum $l-n$.  The subset of states with fixed $n$ form the $n$-th Landau
level which is obviously infinitely degenerate with respect to the angular
momentum.
In a second quantized language the operators $C_l ^n$ satisfy the
usual anticommutation relations
$$ \{ C_{l}^{\dag n} , C_{l'}^{n'} \} = \d _{n,n'} \d _{l,l'} \eqno (2.8)
$$
which, as it is obvious from eq.(2.7) corresponds to the standard
quantization condition
$ \{ \Psi (\vec{x} ,t), \Psi ^{\dag} (\vec{x}',t) \} = \d (\vec{x} - \vec{x}')
$.

So far we have not made a gauge choice. From now on we choose to work in the
symmetric gauge $ \vec{A}= {B \over 2} (y,-x)$. Using the relation$^{[12]}$
$$
\br \vec{x} |n,\z \ke = i^n \sqrt{B \over {2 \pi}} {{(z- \z)^n} \over
\sqrt {n!}} e^{-1/2 |z|^2 + \zbar \z} \eqno (2.9)
$$
where $|n, \z \ke \equiv |n \ke \otimes |\z \ke $ and $z=\sqrt{{ B \over 2}}
(x+iy)$,
we find that the fermion operator is now of the form
$$ \eqalign {
\Psi (\vec{x} ,t) & ={\sqrt{B \over {2 \pi}}} e^{-1/2 |z|^2}  \sum _{n,l =0}
^{\infty} i^n C_l ^n (t)
{{(z-\partial_{\bar{z}})^n} \over {\sqrt{n!}}} {{\bar{z}^{l}} \over
{\sqrt{l!}}} \cr
&= {\sqrt{B \over {2 \pi}}}  e^{-1/2 |z|^2} \sum _{n=0} ^{\infty} \psi ^n
(z, \zbar,t) \cr}
\eqno(2.10)
$$

The quantization condition for the fermion operator constrained to be in the
$n$-th Landau level is
$$ \{\psi^n (z,\zbar,t), \psi^{n \dag}(z',\zbar ',t) \} ={ 1 \over n!}
(z-\del _{\zbar})^n
(\zbar ' - \del _{z'})^n e^{\zbar z'} \eqno (2.11) $$
The rhs of (2.11) is essentially the projection of the $\d$-function
on the $n$-th Landau level.

Let us now consider the case of noninteracting fermions that occupy the first
$n$ Landau levels. In the absence of an external potential the system is
symmetric under independent unitary transformations in the space of $C$'s
acting
at each Landau level, written as:
$$
{C}^I _l (t)= u_{lk} {C}^{I}_k (t)=\br{l} \ve u \ve k \ke {C}^I _k (t) ~~~~~~
I=0,1,...,n-1 \eqno
(2.12)
$$
(We mostly use the notation of ref.[3]) An infinitesimal unitary transformation
is generated by a hermitian operator which we write as
$\ddag \xi ( \hat {b} , {{\hat{b}}^{\dag}} )\ddag$ with the antinormal ordering
symbol, where $\xi$ is a real function when $\hat{b}$ and ${\hat{b}}^{\dag}$
are replaced by $
z$ and $\bar z$ respectively. Using eq.(2.12) we obtain the following
infinitesimal transformation for the $I$-th Landau level fermionic operator:
$$ \eqalign{
\d {\Psi}^I (\vec{x},t) & = i \br \vec{x}| \ddag \xi (b,b^{\dag}) \ddag |I,k
\ke C ^I
_k (t) \cr
& = i \int d^2 \z e^{- | \z |^2} \br \vec{x} |I,\z \ke \br I, \z |
\ddag \xi (b,b^{\dag}) \ddag |I,k \ke C ^I _k (t) \cr
& = i\ddag \xi ( \del_{\zbar}+{z\over 2} , {\zbar \over 2}- \del _z  )\ddag
{\Psi} ^I (\vec{x} ,t) ~~~~~~~~~~I=0,1,...,n-1 \cr} \eqno (2.13)
$$
where $\ddag\ \ \ \ddag$ indicates that the operators $ \del_{\zbar}+{z\over
2}$ act from the left.
The transformations (2.13) preserve the particle
number at each Landau level
$$
\int d {\vec{x}} \d \rho^I (\vec{x},t) =0 ~~~~~~~~~I=0,1,...,n-1 \eqno(2.14) $$
where
$\rho ^I (\vec{x},t)
=  \Psi ^{I \dag}(\vec{x},t) \Psi ^I (\vec{x},t)$ is the $I$-th
Landau level fermion density.
The generators of the transformations (2.13) are given by
$$
\rho^{I} [{\xi}] \equiv
\int  d^2 z e^{-|z|^2} \psi ^{\dag I} (z,\zbar)
\ddag\xi ( \partial_{\bar{z}} , \bar{z}-\del _z )\ddag
\psi ^I (z,\bar{z}) \eqno (2.15) $$
where $d^2 z \equiv {B \over {2 \pi}} dx dy$.
These operators satisfy the commutation rules
$$
[\,\rho ^I [\,\xi_1 \, ],\rho ^J [\, \xi_2 \, ]\, ]= \d ^{IJ}{i\over B}\rho
[\{\!\!\{\xi_1 ,\xi_2 \}\!\!\}] ~~~~~I=0,1,...,n-1 \eqno (2.16)
$$
where
$$
\{\!\!\{\xi_1 ,\xi_2 \}\!\!\}=iB{\sum_{n=1} ^{\infty}}{{(-)^n}\over{n!}}
\left(
{\partial_{z} ^{n}}\xi_1 {\partial_{\bar{z}} ^n}\xi_2 -
{\partial_{\bar{z}} ^{n}}\xi_1 {\partial_{z} ^n}\xi_2\right)\eqno (2.17)
$$
$\{\!\!\{ \}\!\!\}$ is the so-called Moyal bracket. The algebra (2.16)-(2.17)
is the direct sum of $n$ copies of mutually commuting $W _{\infty}$
algebras$^{[13-14]}$. By choosing
$\xi (z,\bar{z})=z^l \bar{z}^k$ we obtain
$$[\, \rho ^I _{rs} \, , \rho ^J _{lk} \,]= \d^{IJ} [\sum_{n=1}^{min(s,l)}
{{(-)^n} \over n!}
{{l!s!} \over {(l-n)!(s-n)!}} \rho_{r+l-n,s+k-n} - (s \leftrightarrow k ,
r \leftrightarrow l)] \eqno(2.18)
$$
where  $\rho ^I _{lk} \equiv \int d^2 z e^{-|z|^2} \psi ^{I \dag} (z, \zbar)
(\del_{\zbar}) ^l (\zbar - \del _z) ^k \psi ^{I} (z, \zbar)$ and it
can be expressed in terms of $C$'s as
$$
\rho ^I _{lk}= \sum _{n=max(0, l-k)} ^{\infty} {{(n+k)!} \over
\sqrt{n!(n+k-l)!}} C^{\dag I} _{n+k-l} C^{I} _{n}   \eqno(2.19)
$$
Let us consider the action of these operators on the ground state $|\Psi
_{\n=n} \ke _0$ , where
$$
| \Psi _{\n =n} \ke _0 = \prod _{I=0} ^{n-1} ( C^{\dag I} _{0} ... C^{\dag I}
_{N-1} ) |0 \ke
\eqno (2.20) $$
for $N' = n N $ electrons.
It is clear that since the operators $\rho ^I _{lk}$ decrease the angular
momentum for $l>k$ and increase the angular momentum for
$l<k$ they satisfy$^{[4]}$
$$ \eqalign{
& \rho^I _{lk} |\Psi_{\n=n} \ke _0=0 ~~~~~~~~~~~~  {\rm if}~~l>k
{}~~~~~~I=0,1,...,n-1 \cr
& \rho^I _{lk} |\Psi_{\n=n} \ke _0=|\Psi_{n,I}> ~~~~~~    {\rm if}~~l\le k
{}~~~~~I=0,1,...,n-1
\cr} \eqno(2.21)
$$
where $|\Psi _{n,I} \ke $ correspond to excitations of higher angular momentum
at
the $I$-th level. The first line in (2.21) can be considered as
the algebraic statement of incompressibility of the $\n =n$ ground state. As
far as excitations are concerned, if $k-l \sim O(1)$, they correspond to
edge excitations of the $I$-th level$^{[15]}$. We
expect that in the semiclassical limit
the \Winf algebra in eqs. (2.16)-(2.17), reduces to the algebra of
area-preserving
diffeomorphisms$^{[3,4,16]}$ (see eq.(2.14)). Upon restriction to the low
energy edge excitations it
gives rise to a $(U(1))^n$ Kac-Moody algebra describing
$n$ independent chiral bosons$^{[15,17-19]}$, one for each Landau level.

\vskip .3 in

\noindent{\bf {3. $W_{\infty}$ algebras in the lowest Landau level}}

In this section we shall consider cases where the
electrons are confined in the lowest Landau level. First we shall briefly
review the derivation of \Winf algebras for
$\n =1/m$ Laughlin states and their relation to $\n =1$ \Winf
algebras, as given in ref.[10], which
will be crucial in constructing similar algebraic structures for quantum Hall
fluids of general filling fraction.

The main point in this derivation is the simple observation that the $\nu =1/m$
Laughlin ground state wavefunction is related to the $\n =1$ wavefunction by
attaching $2p$ (where $m=2p+1$) flux quanta to each electron
$$
\Psi ^0 _{\n =1/m} = \prod _{i<j} (\zbar _{i} - \zbar _{j}) ^{2p} \Psi ^0 _{\n
=1} \eqno (3.1)
$$
In a second quantized language we have that (in this section we have
dropped the superscript $``0"$ used to denote the lowest Landau level fermion
operators)
$$
|\Psi _{\n =1/m} \ke _0 = U_{2p} |\Psi  _1 \ke _0 ,  ~~~~~~~~
|\Psi  _1 \ke _0  = \tilde{U} _{2p} |\Psi  _{\n=1/m} \ke _0  \eqno (3.2)
$$
where
$$ \eqalign{
& |\Psi_1 \ke _0= \int d^2 z_1...d^2 z_N e^{-\sum_{i}
|z_i|^2} \prod_{i<j}
(\bar{z}_i-\bar{z}_j) \kvecz \cr
& |\Psi _{\n=1/m} \ke _0= \int d^2 z_1...d^2 z_N
e^{-\sum_{i} |z_i |^2}  \prod_{i<j}(\zbar _i -\zbar _j )^m \kvecz \cr}
\eqno (3.3)$$
and
$$\eqalign{
U_{2p}&= \sum _{N=2} ^{\infty} \int d^2 z_1...d^2 z_N e^{-\sum |z_i|^2}
\prod_{i<j}
(\bar{z}_i-\bar{z}_j)^{2p} |z_1...z_N \ke \br z_1...z_N| \cr
\tilde{U}_{2p} &= \sum _{N=2} ^{\infty} \int d^2 z_1...d^2 z_N e^{-\sum |z_i|^2
} \prod_{i<j}
(\bar{z}_i-\bar{z}_j)^{-2p} |z_1...z_N \ke \br z_1...z_N| \cr} \eqno(3.4)
$$
where $|z_1...z_N \ke ={1 \over
\sqrt{N!}}\psi^{\dag}(z_1)...\psi^{\dag}(z_N)|0 \ke $.
The operator $\tilde{U} _{2p}$ is in general singular but its action on
the space of the $\n =1/m$ Laughlin states is well defined and
$\tilde{U} _{2p} U_{2p} = U_{2p} \tilde{U} _{2p} = 1$.
It is clear now that
$$W_{2p}[\xi] \equiv U_{2p}~\rho[\xi] ~\tilde{U}_{2p}  \eqno(3.5) $$
are the corresponding \Winf algebra generators for $\n =1/m$. They are mainly
the
original $\n =1$ generators transformed by a similarity
transformation.{\footnote{*}{Such transformations have been recently used in
the context of quantum gravity in ref.[27].}}
Corresponding to (3.5) there is a second quantized
expression in terms of fermion and quasihole operators
$$W_{2p}[\xi]= \int d^2 z e^{-|z|^2} \psi^{\dag} (z) e^{2p \alpha
(\bar{z})} \ddag \xi ( \partial _{\bar{z}}, \bar{z}) \ddag e^{-2p \alpha
(\bar{z})} \psi(\bar{z}) \eqno(3.6)
$$
where
$$\alpha (\bar{z})=\int d^2 z' e^{-|z'|^2} \ln (\bar{z}-\bar{z}')
\psi^{\dag}(z') \psi (\bar{z}') \eqno (3.7) $$
and $e^{\a (\zbar)}$ is the quasihole operator since
$$
e^{\alpha (\zbar )}| \Psi \ke = \int d^2 z_1...d^2 z_N
e^{-\sum_{i}|z_i |^2}  F(\zbar _1,...,\zbar _N)
\prod_i (\zbar -\zbar _i )\kvecz \eqno (3.8)
$$
where
$$
| \Psi \ke = \int d^2 z_1...d^2 z_N
e^{-\sum_{i}|z_i |^2} F(\zbar _1,...,\zbar _N) \kvecz \eqno (3.9)
$$
We see that the appropriate insertion of the vortex (quasihole) operators in
(3.6) reflects the underlying similarity transformation (3.5). Using the
fact that
$$ \eqalignno {
 \psi(\bar{z}') e^{n \alpha (\bar{z})} & = (\bar{z}\
-\bar{z}')^{n} e^{n \alpha (\bar{z})} \psi(\bar{z}') \cr
e^{n \alpha(\bar{z})}\psi^{\dag} (z')  & = (\bar{z} -
\partial_{z'})^{n} \psi ^{\dag} (z') e^{n \alpha (\bar{z})}
&(3.10) \cr}
$$
where $n$ is an integer and $[ \a (\zbar), \a (\zbar ')] =0$
we can show that the operators $W_{2p}$ satisfy a strong \Winf algebra
$$[W_{2p}[\xi _1], W_{2p}[\xi _2]] = W_{2p}[\{\!\!\{\xi _1 ,
\xi _2 \}\!\!\}] \eqno(3.11)
$$
where $\{\!\!\{ \}\!\!\}$ is the Moyal bracket defined earlier in (2.17).
They further play the role of a spectrum generating algebra{\footnote {*} {
We have used the infinite plane geometry where the electrons are confined by
the existence of an external confining potential$^{[4,10]}$, for example
a central harmonic oscillator potential. Such a potential can be thought of
controlling the maximum
single-particle angular momentum $L$ available within a Landau level, which
also determines the size of the droplet corresponding to the ground state
configuration. In the case of the $\n =1/m$ ground state, eqs.(3.3)-(3.13),
$N$ and $L$ are related by $L=m(N-1)$.}}
in the space of Laughlin states, since
$$ \eqalignno{
& (W_{2p})_{lk} |\Psi_{\n=1/m} \ke _0=0 ~~~~~~~~~~~~  {\rm if}~~l>k \cr
& (W_{2p})_{lk} |\Psi_{\n=1/m} \ke _0 =|\Psi_{\n=1/m} \ke ~~~~~~    {\rm
if}~~l\le k
&(3.12) \cr}
$$
where $|\Psi _{\n =1/m} \ke $ is a higher angular momentum state of the form
$$ |\Psi_{\n=1/m} \ke = \int d^2 z_1...d^2 z_N e^{-\sum_{i}
|z_i|^2} \prod_{i<j}
(\bar{z}_{i}-\bar{z}_{j})^m P(\bar{z}_1,...,\bar{z}_N)
|z_1...z_N \ke \eqno (3.13)$$
and $P(\bar{z}_1,...,\bar{z}_N)$ is a homogeneous symmetric polynomial.

The operators $W_{2p}$, $p=0,1,...$, form a one-parameter
family of representations for $W_{\infty}$ algebra and the corresponding
Laughlin ground state is the highest weight state.
This provides an algebraic statement of incompressibility for the Laughlin
ground states.

We would like now to extend these ideas to include other incompressible
states corresponding to filling fractions $\n \ne 1/m$. We still consider the
case of lowest Landau level fermions. First we identify the \Winf
algebra structure corresponding to the state with filling fraction $\n =1 -
1/m$. Using the idea of particle-hole conjugation$^{[20]}$, we can write the
$\n =1 -{1 \over m}$ ground state, in the thermodynamic limit and up to
normalization factors, as
$$
|\Psi _{\n =1 - 1/m} \ke _0 \sim  \int d^2 z_1 ... d^2 z_M e^{-\sum |z_i|^2}
\prod
_{i<j} (z_i
- z_j) ^m \psi (\zbar _1) ... \psi (\zbar _M) |\Psi _{\n =1} \ke _0 \eqno(3.14)
$$
Let us now introduce the operator $\b (z)$
$$\b (z)=\int d^2 z' e^{-|z'|^2} \ln (z-z')
\psi (\zbar ') \psi ^{\dag} (z') \eqno(3.15) $$
The operator $e^{\b (z)}$ satisfies equations similar to (3.10):
$$ \eqalign {
e^{n \b (z)} \psi(\bar{z}')  & = (z -
\partial_{\zbar '})^{n}   \psi(\bar{z}') e^{n \b (z)} \cr
\psi^{\dag} (z') e^{n \b (z)}  & = (z
-z')^{n} e^{n \b (z)} \psi ^{\dag} (z')
\cr} \eqno(3.16)
$$
Based on these relations and the fact that $\psi ^{\dag} (z_i) |\Psi _{\n
=1} \ke _0 =0 $, we find that the operators
$$
\tilde{W} _{2p}[\xi]= \int d^2 z e^{-|z|^2} \psi (\zbar) e^{2p \b
(z)} \ddag \xi ( \partial _{z}, z) \ddag e^{-2p \b
(z)} \psi ^{\dag} (z) \eqno(3.17)
$$
satisfy a \Winf algebra and the state (3.14) satisfies a highest weight
condition
$$
(\tilde{W} _{2p})_{lk} |\Psi _{\n = 1- 1/m} \ke _0 =0 ,~~~~~~~~~~l>k
\eqno(3.18)
$$
The operator $\tilde{W} _{2p}$ is essentially the charge-conjugated version of
$W _{2p}$.

In order to obtain more general filling fractions, one may consider systems
where two distinct species of electrons are involved (cases where the electrons
have different spin or they are in separate layers). Trial ground state
wavefunctions of the form
$$
\Psi ^{m_1,m_2,n}  (\vec{x}_i, \vec{\w}_i) = \prod _{i<j} (\zbar _i  - \zbar _j
)^{m_1}
\prod_{i<j}  (\wbar _i  - \wbar _j ) ^{m_2} \prod _{i,j} (\zbar _i - \wbar
_j)^{n} e^{-1/2 \sum _{i} (|z_i|^2 + |w_i|^2)}
\eqno(3.19) $$
have been suggested$^{[21]}$ as candidates for describing incompressible Hall
states for
a two-layer system at $\n = {{m_1 +m_2 -2n} \over {m_1 m_2 - n^2}}$. The
integers
$m_1$ and $m_2$ are odd so that the wavefunctions are antisymmetric under
exchange of identical fermions.

In order to identify the quantum \Winf algebra structure associated with
(3.19) we introduce two independent lowest Landau level fermion operators such
that
$$ \eqalign{
\{ \psi ^I (\zbar,t), \psi ^{\dag J} (z',t) \} = \d ^{IJ} e^{\zbar z'}
{}~~~~~~~~I=1,2 \cr
\{ \psi ^I (\zbar,t), \psi ^J (\zbar', t) \} = \{ \psi ^{\dag I} (z, t),
\psi ^{\dag J} (z',t) \} = 0 \cr}
\eqno (3.20)
$$
and the corresponding quasihole operators $e^{\a ^{I} (\zbar)}$ where
$$\alpha ^I (\bar{z})=\int d^2 z' e^{-|z'|^2} \ln (\bar{z}-\bar{z}')
\psi^{\dag I}(z') \psi ^I (\bar{z}') \eqno (3.21) $$

A relation similar to (3.1) applies between (3.19) and the $\n=1$ state. The
main difference
now is that vortices from different species are involved. This suggests that
the corresponding \Winf generators have the form
$$ \eqalign {
W^1 [\xi] & = \int d^2 z e^{-|z|^2} \psi^{1 \dag} (z)
e^{ [(m_1 -1) \a ^1 (\zbar) + n \a ^2
(\bar{z})]} \ddag \xi ( \partial _{\bar{z}}, \bar{z}) \ddag e^{- [(m_1 -1)
\a ^1 (\zbar) + n \a ^2
(\bar{z})]}
 \psi ^1 (\bar{z}) \cr
W^2 [\xi] & = \int d^2 z e^{-|z|^2} \psi^{2 \dag} (z)
e^{ [( n \a ^1 (\zbar) + (m_2 - 1) \a ^2
(\bar{z})]} \ddag \xi ( \partial _{\bar{z}}, \bar{z}) \ddag e^{- [n
\a ^1 (\zbar) + (m_2 -1) \a ^2
(\bar{z})]}
 \psi ^2 (\bar{z}) \cr}
\eqno(3.22)
$$
It is straightforward to show that the operators in (3.22) give rise to two
commuting \Winf algebras and that the corresponding ground state
(3.19) satisfies highest weight condition. The expressions (3.22) provide
a family of representations of \Winf parametrized by three integers.

Using the expressions (3.21) for the quasihole operators we can also construct
second quantized operators for the order parameters $^{[22,23]}$ of the system
$$ \eqalign{
& q^{\dag} _1 = \int d^2 z e^{-|z|^2} \psi ^{1 \dag} (z) e^{m_1 \a ^1 (\zbar)}
e^{n \a ^2 (\zbar)} \cr
& q^{ \dag} _2 = \int d^2 z e^{-|z|^2} \psi ^{2 \dag} (z) e^{m_2 \a ^2 (\zbar)}
e^{n \a ^1 (\zbar)} \cr}
\eqno (3.23)
$$
This is a straightforward generalization of the order parameter operator for
the $\n = 1/m$ Laughlin state$^{[10]}$.

Using eqs.(3.10) and the commutativity of $\a$'s we can show that
$$ \eqalign{
& [q ^{\dag} _1, q ^{\dag} _1] = [q ^{\dag} _2, q ^{\dag} _2 ]=0 \cr
& q ^{\dag} _1 q ^{\dag} _2= - (-) ^n q^{\dag} _2 q ^{\dag} _1 \cr}
\eqno (3.24)
$$
Thus for $n$ odd the operators $q^{\dag} _I$ are bosonic, as there should be in
order to describe order parameters. For $n$ even they are fermionic and so we
shall consider the redefined operators
$$
Q^{\dag} _I = q^{\dag} _I  e^{i {\pi \over 2} S}, ~~~~~~ S= \int d^2 z
e^{-|z|^2}( \psi ^{1 \dag}(z) \psi ^1 (\zbar) - \psi ^{2 \dag}(z) \psi ^2
(\zbar))
\eqno (3.25)
$$
where $S$
is a ``cocycle" factor needed to impose commutativity conditions$^{[23]}$ for
$Q^{\dag} _1$, $Q^{\dag} _2$
$$
[Q^{\dag} _I, Q^{\dag} _J] =0 , ~~~~~~~~I,J=1,2 \eqno (3.26)
$$
The many-body ground state (3.19) is now created out of the vacuum by the
action
of the bosonic operators
$$
|\Psi ^{m_1,m_2,n} \ke = {1 \over \sqrt{N!M!}} (Q^{\dag} _2)^M (Q^{\dag} _1)^N
|0 \ke
\eqno (3.27)
$$

The above construction can now be generalized to an $r$-layer system. The
generalized ground state wavefunction is$^{[24]}$
$$
\Psi ^K (\vec{x} _i ^I) = \prod _{I=1}^{r} \prod _{i<j} (\zbar _i ^I -
\zbar _j ^I)
^{K_{II}} \prod_{I<J} \prod _{i,j} (\zbar _i ^I - \zbar _j ^J) ^{K_{IJ}}
e^{-1/2 \sum _{iI} |z_i|^2}
\eqno (3.28)
$$
where $K_{II}$ are odd integers . The corresponding
filling fraction is given by  $\n = \sum _{I,J} (K^{-1})_{IJ}$ where $K$ is an
$r \times r$ symmetric matrix.
The \Winf generators are
$$
W_{\tilde{K}} ^I [\xi]= \int d^2 z e^{-|z|^2} \psi^{\dag I} (z)
e^{ \sum _{J} \tilde{K}_{IJ} \alpha ^J
(\bar{z})} \ddag \xi ( \partial _{\bar{z}}, \bar{z}) \ddag e^{- \sum _{j}
\tilde {K} _{IJ} \alpha ^J
(\bar{z})} \psi ^I (\bar{z}) \eqno(3.29)
$$
where $I=1,...,r$ and $\tilde {K} = K -{{\rm I}\!{\rm I}}$
(${{\rm I}\!{\rm I}}$ is the identity matrix). We can show, as before,
that the operators $W ^I _{\tilde{K}}$ give rise to $r$ commuting copies of
\Winf algebras and the ground states corresponding to (3.28)
satisfy highest weight conditions. Expressions (3.29) provide a
family of representations of \Winf algebra parametrized by the symmetric matrix
$K$.

As far as the order parameters are concerned we can appropriately generalize
the
expressions (3.25)-(3.27). We start by noticing that
$$
q^{\dag} _I  q^{\dag} _J = -(-)^{K_{IJ}} q^{\dag} _J q^{\dag} _I \eqno (3.30)
$$
where
$$
q ^{\dag} _I = \int d^2 z e^{-|z|^2} \psi ^{I \dag} (z) e^{\sum _{L} K_{IL} \a
^{L} (\zbar)}
\eqno (3.31)
$$
We then find that the appropriate bosonic order parameter operators are
$$
Q^{I \dag} = q^{I \dag} e^{i {\pi \over 2} S},~~~~~~
S=  \sum _{I} ~ \L _I ~ \int d^2 z e^{-|z|^2} \psi ^{I \dag} (z) \psi ^I
(\zbar)
\eqno (3.32)
$$
where $\L _I$ are integers such that
$ K_{IJ} + {{\L _I - \L _{J}} \over 2} = {\rm odd} $.

As we noticed in ref.[10] the \Winf generators, eq.(3.29), can be
further written as
$$
W_{\tilde{K}}^I [\xi]= \int d^2 z e^{-|z|^2} \psi^{\dag I} (z) \ddag \xi
(\partial _{\bar{z}}-\sum _{J} \tilde{K}_{IJ} \int d^2 z' {{\rho ^{J} (z',
\bar{z}')}
\over {(\bar{z}-\bar{z}')}}, \bar{z}) \ddag \psi ^{I}(\bar{z}) \eqno(3.33)
$$
The term $ \sum_{J} \tilde{K} _{IJ} \int d^2 z' {{\rho ^{J} (z',\bar{z}')}
\over
{(\bar{z}-\bar{z}')}}$ plays the role of a gauge potential and it is similar
to the one induced by the Chern-Simons interaction$^{[25]}$. Analogous first
quantized
expressions were also derived in ref.[9].

Further by applying particle-hole conjugation as in (3.14)-(3.18) we can
derive second quantized expressions for the \Winf generators for
$\tilde{\n} = 1-\n$ by charge-conjugating $W_{\tilde{K}} ^I$.

The wavefunctions (3.28) are also associated to the standard hierarchy
scheme$^{[6,7,24]}$.
They are written in terms of electron coordinates and the coordinates of
the quasiparticles formed at each level of the hierarchy. According to the
hierarchy scenario, the quasiparticles at each level are the condensates of the
quasiparticles of the previous level. Although it is clear from eq.(3.33) that
at a first quantized level there is a \Winf structure$^{[9]}$, the precise
second quantized
representation of \Winf generators is not clear. The main difficulty in this
case is the assignment of creation and annihilation operators to the
quasiparticles at each level of the hierarchy. Given this difficulty we are
going to use in the next section, the alternative description of the FQHE,
proposed by Jain, where everything is formulated in terms of electrons. Such a
framework will be very convenient in displaying the \Winf algebra structure
corresponding to the FQHE states, as described by Jain wavefunctions.

\vskip .3in
\noindent{\bf 4. {\Winf algebras for Jain states}}

One of the theories proposed to explain the observed fractions for the FQHE is
the one suggested by Jain$^{[8]}$, in which there is a strong connection
between
IQHE and FQHE. As far as the hierarchical filling fractions $\n= {n \over
{2pn+1}}$ are concerned, the essential idea of this scenario is that the
corresponding incompressible FQHE ground state wavefunction $\Psi _{\n} ^0$, is
related to the IQHE wavefunction $\Psi _n ^0$ as
$$ \Psi _{\n} ^0 = \prod _{i<j} (\zbar _i - \zbar _j)^{2p} \Psi _n ^0 \eqno
(4.1)
$$
This relation is a straightforward generalization of (3.1) and we can therefore
use previous ideas in order to write down the second quantized \Winf generators
which would express the incompressibility of (4.1). The main difference now
is that fermions in higher Landau
levels are involved and some of the previously used operator relations
might change.

The state corresponding to (4.1) can be written as ($N' =n N$)
{\footnote{$\dag$} {$N$ is related to the maximal single-particle angular
momentum available within a Landau level as it was mentioned in the footnote in
pg.8. For example by counting powers of $\zbar ^0 _i$ we find that $2p(N'-1) +
(N-1) =L$, where $L$ is the maximal single-particle angular momentum in the
lowest Landau level.}}
$$ \eqalign {
|\Psi _{\n} \ke _{0} \sim & \int \prod _{I=0} ^{n-1} \left[ (\prod _{i=1} ^{N}
d^2 z_i ^I
e^{-|z_i^I|^2}) \prod _{i<j} (\zbar _i^I - \zbar _j^I) ^{2p} \right]
\prod _{I<J} \prod_{i,j} (\zbar _i^I
-\zbar_j^J)^{2p} \cr
& \prod_{I=0}^{n-1} \left[ \D ^I
[\vec{x}_1^I,...,\vec{x} _N^I] \psi ^{\dag I} (z_1^I, \zbar _1^I)...\psi ^
{\dag I}
(z_N^I, \zbar _N^I) \right] |0 \ke \cr}
\eqno (4.2)
$$
where $\D ^I [\vec{x}_1,...,\vec{x} _N]$ {\footnote{*} {In defining
$\D ^I [\vec{x}_1,...,\vec{x} _N]$ we have factored out all the exponential
factors $e^{-{1 \over 2} |z_i|^2}$. For example for the lowest Landau level
$\D ^0 [\vec{x}_1,...,\vec{x} _N] = \prod _{i<j} (\zbar _i - \zbar _j)$.}}
is the properly antisymmetrized
$N$-body wavefunction and $\psi ^I (z, \zbar)$ is the fermion operator at the
$I$-th Landau level. Antisymmetrization is done
independently at each Landau level. We first notice that the incompressibility
condition for $|\Psi _{\n} \ke _{0}$ in the first line of (2.21) can be
expressed as a differential equation on $\D$, namely
$$
\sum _{i=1} ^N (\del {\zbar _i})^l ( \zbar _i - \del _{z_i})^k \D ^I
[\vec{x}_1,...,\vec{x}_N] =0  ~~~~~~~~~~~~~l>k, ~~~~~\forall I  \eqno (4.3)
$$

Let us again consider the operators $\a ^{I} (\zbar)$, $I=0,1,...,n-1$.
$$\alpha ^{I} (\bar{z})=\int d^2 z' e^{-|z'|^2} \ln (\bar{z}-\bar{z}')
\psi ^{\dag I} (z', \zbar ') \psi ^I (z', \zbar ') \eqno (4.4) $$
Using the commutation relations (2.11) we
can show that
$$ \eqalign{
& [\a ^I (\zbar),~ \psi ^{\dag I}
(z', \zbar ')] =  \ln (\zbar - \del _{z '})
\psi ^{\dag I} (z', \zbar ') \cr
& [\a ^I (\zbar),~ \psi ^I
(z', \zbar ')] =  - \ln (\zbar - \zbar ' + \del _{z '})
\psi ^{I} (z', \zbar ') \cr} \eqno (4.5)
$$
Using these relations along with the fact that wavefunctions in different
Landau levels are orthogonal we find that the operators
$$
W_{2p} ^{I} = \int d^2 z e^{-|z|^2} \psi ^{\dag I} (z, \zbar) e^{2p \sum _J \a
^J (\zbar)} \ddag \xi (\del _{\zbar}, \zbar - \del _z)
\ddag e^{-2p \sum _J \a ^J (\zbar)} \psi ^I (z, \zbar) \eqno (4.6)
$$
where $I=0,...,n-1$, give rise to $n$ commuting copies of \Winf algebras and
the corresponding Jain state for $\n ={n
\over {2pn+1}}$ satisfies the highest weight condition
$$
(W^I _{2p})_{lk} |\Psi _{\n} \ke _{0} =0 ,~~~~~~~~~~~~l>k ~~~~~~I=0,...,n-1
\eqno (4.7)
$$
As before we can use particle-hole conjugation to derive the charge-conjugated
$\tilde{W} ^I _{2p}$ operators which express the incompressibility of the
states
with $\tilde{\n} = 1- \n$.

\vskip 0.3 in
\noindent {\bf 5. Discussion}

In this paper we have given an explicit second quantized representation of the
\Winf algebra structure characterizing the incompressibility of a large class
of fractional Hall states. The corresponding generators are expressed in terms
of fermion and vortex operators. In general incompressible states can be
thought of as highest weight states of \Winf algebra$^{[11]}$. Expressing the
\Winf generators in terms of Fock operators is useful in constructing
representations and of course the highest weight state. It is thus possible to
obtain new candidate states for describing quantum Hall effect.

For the specific examples we considered, the multilayer
incompressible states and the Jain states, we find that there are $r$
commuting copies of \Winf algebras, where
$r$ is the number of layers in a multilayer system or the number of Landau
levels in the Jain framework$^{[8]}$. This \Winf algebra, besides
characterizing the
incompressibility of the ground state, provides a spectrum generating algebra
for excitations. We expect
that in the semiclassical limit the \Winf algebra reduces to the algebra of
area-preserving diffeomorphisms, which upon restriction to the low energy
edge excitations gives rise to a $((U(1)) ^r$
Kac-Moody algebra describing $r$ independent chiral bosons, one for each
layer. So far there is the underlying assumption that the number of electrons
at each layer is conserved. If this is relaxed by considering tunneling between
different layers the spectrum generating algebra is enlarged by an embedding in
a nonabelian structure.{\footnote {*}{This was pointed to me by Bunji Sakita.}
}

Further in constructing the \Winf generators for multilayer systems and the
hierarchical states based on Jain's construction we made use of the special
mapping (4.1) between integer and fractional quantum Hall states. In Jain's
scenario there are more generalized incompressible states$^{[8]}$, for example
states
which can be written as products of integer quantum Hall states. A generic
feature of the corresponding incompressible fluids  is the existence of
nonabelian edge excitations$^{[26]}$, which again suggests a nonabelian \Winf
algebra
structure. The precise fermionic representation of this is under investigation.

An approach somewhat similar in spirit, in linking the incompressibility of
fractional quantum Hall states to the existence of a \Winf algebra structure,
has been advocated in ref.[11]. Using the representation theory of a centrally
extended \Winf algebra (corresponding to the one-dimensional edge degrees of
freedom) the authors of [11] were able to classify the hierarchical (abelian)
quantum Hall states. Although the relation between the two-dimensional
centerless \Winf algebra and the one-dimensional centrally extended one is
straightforward in the $\n =1$ case, it is not so for other filling
fractions. In the $\n =1$ case, the normal ordering of the full two-dimensional
\Winf algebra with respect to the $\n =1$ ground state gives the
centrally extended one-dimensional \Winf algebra$^{[19]}$. It is interesting
to see whether the two \Winf structures can be
similarly related in the case of other filling fractions.

\vskip 0.3 in
\noindent{\bf Acknowledgements}

I would like to thank T.R. Govindarajan, V.P. Nair and B. Sakita for useful
discussions. This work was supported by the U.S. Department of Energy under the
contract
number DE-FG02-85ER40231.

\vskip 0.3 in
\centerline{\bf References}

\item{[1]} R.E. Prange and S.M. Girvin, {\it ``The Quantum Hall Effect''},
Springer, New York, 1990.
\item{[2]} M. Stone, {\it ``Quantum Hall Effect''}, World Scientific, 1992.
\item{[3]} S. Iso, D. Karabali and B. Sakita, {\it Phys. Lett.} {\bf B296}
(1992) 143.
\item{[4]} A. Capelli, C. Trugenberger and G. Zemba, {\it Nucl. Phys.}
{\bf B396} (1993) 465.
\item{[5]} R.B. Laughlin in ref.[1]; {\it Phys. Rev. Lett.} {\bf 50} (1983)
1395.
\item{[6]} F.D.M. Haldane in ref. [1].
\item{[7]} F.D.M. Haldane, {\it Phys. Rev. Lett.} {\bf 51} (1983) 605; B.I.
Halperin, {\it Phys. Rev. Lett.} {\bf 52} (1984) 1583.
\item{[8]} J.K. Jain, {\it Phys. Rev. Lett.} {\bf 63} (1989) 199; {\it Phys.
Rev.} {\bf B40} (1989) 8079; {\it Phys. Rev.} {\bf B41} (1990) 8449.
\item{[9]} M. Flohr and R. Varnhagen, BONN-HE-93-29 preprint, hep-th/9309083.
\item{[10]} D. Karabali, SU-4240-553 preprint, cond-mat/9309050, to appear in
{\it Nucl. Phys.} {\bf B}.
\item{[11]} A. Cappelli, C. Trugenberger and G. Zemba, {\it Phys. Rev. Lett.}
{\bf 72} (1994) 1902.
\item{[12]} B. Sakita in {\it Proceedings, 16th Taniguchi Int. Symposium on the
theory of condensed matter}, eds. N. Kawakami and A. Okiji, Springer Verlag (to
be published).
\item{[13]} I. Bakas, {\it Phys. Lett.} {\bf
B228} (1989) 57; I. Bakas and E. Kiritsis {\it Nucl. Phys.} {\bf B343} (1990)
185.
\item{[14]} D.B. Fairlie and C.K. Zachos, {\it Phys. Lett.}
{\bf B224} (1989) 101; J. Hoppe and P. Schaller,
{\it
Phys. Lett.} {\bf B237} (1990) 407; C.N. Pope, L.J. Romans and X. Shen {\it "A
Brief History of $W_{\infty}$''} in Strings 90, ed. Arnowitt et al (World
Scientific 1991) and references therein.
\item{[15]} for a review on edge excitations see, X.G. Wen, {\it Int. J. Mod.
Phys.} {\bf B6} (1992) 1711.
\item{[16]} A. Cappelli, C.A. Trugenberger and G.R. Zemba, {\it Phys. Lett.}
{\bf
B306} (1993) 100.
\item{[17]} M. Stone, {\it Phys. Rev.} {\bf B42} (1990) 8399; {\it Ann. Phys.}
{\bf 207} (1991) 38.
\item{[18]} S. Iso, D. Karabali and B. Sakita, {\it Nucl. Phys.} {\bf B388}
(1992) 700.
\item{[19]} A. Cappelli, G.V. Dunne, C.A. Trugenberger and G.R. Zemba, {\it
Nucl.
Phys.} {\bf B398} (1993) 531; B. Sakita, {\it Phys. Lett.} {\bf B315} (1993)
124.
\item{[20]} S.M. Girvin, {\it Phys. Rev.} {\bf B29} (1984) 155.
\item{[21]} B.I. Halperin, {\it Helv. Phys. Acta} {\bf 56} (1983) 75; E.H.
Reyazi
and H.D.M. Haldane, {\it Bull. Am. Phys. Soc.} {\bf 32} (1987) 892; D.
Yoshioka, A.H.
MacDonald and S.M. Girvin, {\it Phys. Rev.} {\bf B39} (1989) 1932; F. Wilczek,
{\it Phys. Rev. Lett.} {\bf 69} (1992) 132.
\item{[22]} N. Read, {\it Phys. Rev. Lett.} {\bf 62} (1989) 315.
\item{[23]} G. Moore and N. Read, {\it Nucl. Phys.} {\bf B360} (1991) 362.
\item{[24]} N. Read, {\it Phys. Rev. Lett.} {\bf 65} (1990) 1502; J. Frolich
and A. Zee, {\it Nucl. Phys.} {\bf B354} (1991) 369; X.G.
Wen and A.Zee, {\it Phys. Rev.} {\bf B46} (1992) 2290.
\item{[25]} A. Lopez and E. Fradkin, {\it Phys. Rev.} {\bf B44} (1991) 5246.
\item{[26]} X.G. Wen, {\it Phys. Rev. Lett.} {\bf 66} (1991) 802; B. Blok and
X.G. Wen, {\it Nucl. Phys.} {\bf B374} (1992) 615.
\item{[27]} J. Ellis, N.E. Mavromatos and D.V. Nanopoulos, CERN-TH 7195/94,
ENS-LAPP-A-463/94, ACT-5/94, CTP-TAMU-13/94 preprint.

\end